\documentclass[twocolumn,showpacs,amsmath,amssymb,floatfix,aps]{revtex4}
\usepackage{graphicx}

\begin{document}

\title{Background Independent Quantum Field Theory\\ and the Cosmological Constant Problem}
\date{\today}
    \author{Olaf Dreyer}
    \email{odreyer@perimeterinstitute.ca}
    \affiliation{Perimeter Institute for Theoretical Physics, 35 King
    Street North, Waterloo, Ontario N2J 2W9, Canada}

\begin{abstract} We introduce the notion of background independent quantum field theory. The distinguishing feature of this theory is that the dynamics can be formulated without recourse to a background metric structure. We show in a simple model how the metric properties of spacetime can be recovered from the dynamics. Background independence is not only conceptually desirable but allows for the resolution of a problem haunting ordinary quantum field theory: the cosmological constant problem. 
\end{abstract}


\maketitle

\section{Introduction}
There are currently two competing views of the role quantum field theory plays in our theoretical understanding of nature. In one view, quantum field theory describes the dynamics of the elementary constituents of matter. The job of the physicist is to figure out what the elementary particles are and to find the appropriate Lagrangian that describes the interactions. The Standard Model of elementary particle physics is the epitome of this view (see \cite{qft} for an authoritative exposition of this view). The other point of view likens the use of quantum field theory to its use in other areas of physics, most importantly in solid state physics. Here, quantum field theory is not used to describe the dynamics of elementary particles. In solid state physics, this would be fruitless, since the dynamics of a large number of atoms is usually beyond anyone's ability to compute. It turns out, however, that these large numbers of constituents often have collective excitations that can be well-described by quantum field theory and that are responsible for the physical properties of the material. The view is then that the elementary particles of the Standard Model are like the collective excitations of solid state physics. The world we live in is just another material whose excitations are described by the Standard Model. The point of view was introduced by P.~W.~Anderson \cite{anderson} and has since found a large following (see e.g. \cite{volovik, laughlin, wen} and references therein). 

The search for a quantum theory of gravity requires a unification of quantum field theory and general relativity. If the second point of view is correct, it should not be restricted to the Standard Model, but also include gravity.  From this perspective, the second point of view possesses one objectionable feature:  the material 
whose excitations give rise to the elementary particles around us is assumed to rest in a Newtonian world whose notions of distance are taken over to formulate the theory.  That is, the condensed matter models are manifestly \emph{background-dependent}.  Since general relativity is background-independent, namely, space and time are dynamical degrees of freedom, it is difficult to see how gravity can be included in this second point of view. 

In this paper, we want to address this objection by revisiting the solid state models from a background independent perspective. One can formulate the dynamics of a large class of models without recourse to a background structure. Notions of distance can then be recovered from the dynamics of the theory. 

The lack of background independence is not just a conceptual shortcoming. We argue that if the theory is formulated in a background independent way it allows for the solution  of one of the long-standing problems of quantum field theory: the cosmological constant problem. 

This problem arises when one views quantum field theory as a theory describing fields living on a curved spacetime. This view runs into a serious problem when one considers the effect the quantum fields should have on spacetime. Since all the modes of the quantum field have a zero energy of $\pm 1/2\hbar\omega$\footnote{The sign depends on whether the particles are fermions or bosons.}, one expects a contribution to the vacuum energy on the order of 
\begin{equation}
\int^\Xi d\omega\;\hbar\omega^3 \sim \hbar\,\Xi^4,
\end{equation}
where $\Xi$ is some high energy cut-off. If one takes this cut-off to be the Planck energy the vacuum energy is some 123 orders of magnitude away from the observed value of the cosmological constant \cite{wmap}, making this the worst prediction in theoretical physics (for detailed discussions of this problem see \cite{weinberg,ccp}).

We will see that the cosmological constant problem arises \emph{only} when one regards the quantum fields as living on the background. This is where this problem connects with the conceptual problem of background dependence. If, instead, it is the quantum fields that make the spacetime appear in the first place, and they are not treated as living on the background, then the cosmological constant problem disappears. 

The organization of the paper is as follows. In the next section we demonstrate what we mean by background independent quantum field theory in a very simple model, in which we show how the dynamics of the model can be formulated without a given background structure. We indicate how the spacetime picture can be recovered from the dynamics of the theory. Section \ref{sec:ccp} contains the argument why the cosmological constant problem does not appear in background independent quantum field theory. To show this we look at the coarse grained description of the model in section \ref{sec:model}. We show that where there was no cosmological constant problem in the original model, it appears if a background is assumed. In the concluding section \ref{sec:conclusion}, we remark on the role special relativity plays in our approach and discuss what our results mean for the search for a theory of quantum gravity.

\section{ A Model}\label{sec:model}
We will illustrate background independent quantum field theory on a very simple example: The XY-model in one spatial dimension. The Hamiltonian is given by
\begin{equation}
H = J_\perp\sum_{i=1}^N (S^+_iS^-_{i+1} + S^-_iS^+_{i+1}).
\end{equation}
The operators $S^\pm$ are formed from the Pauli matrices $\sigma^a$, $a=1,2,3$:
\begin{eqnarray}
S^\pm & = & S^1 \pm i S^2, \\
S^a & = &  \frac{1}{2}\sigma^a.
\end{eqnarray}
The index $i$ ranges over the $N$ lattice sites of the one-dimensional lattice. We have assumed only nearest neighbor coupling and have not included the interaction between the $3$-components of the spins. We are thus in the quantum limit of the one-dimensional spin chain. We will impose a circle topology, so that $N+1$ is to be identified with 1. 

\begin{figure}
\begin{center}
\includegraphics[height=3.5cm]{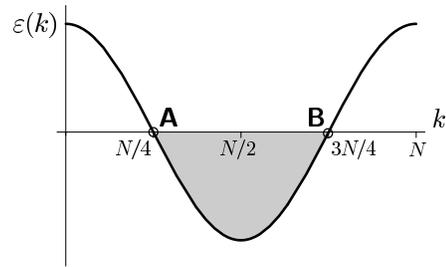}
\end{center}
\caption{The dispersion relation for the one-dimensional XY-model. In the state where all the negative energy levels are filled the dispersion relation becomes linear at the points \textbf{\textsf{A}} and \textbf{\textsf{B}}. Close to these points the spin chain looks relativistic.}\label{fig:spectrum}
\end{figure}

The $S^\pm$ satisfy mixed commutation relations. They commute for different sites and anti-commute on the same site:
\begin{eqnarray}
\{S^\pm_i, S_i^-\} & = & 1, \\
 & & \nonumber \\
{[}S^+_i, S^-_j{]} & = & 0, \text{ for } i\ne j.
\end{eqnarray}
The $S^\pm$ can thus be thought of as creating or annihilating hard-core bosons. To have more standard commutation relations, and also to more readily solve the system, one can perform a Jordan-Wigner transformation \cite{jordanwigner} to purely fermionic variables, $f_i$, $i = 1,\ldots,N$:
\begin{equation}\label{eqn:fermions}
f_i = U_iS_i^-,  \ \ \ f_i^\dagger = S_i^+ U_i^\dagger
\end{equation}
\begin{equation}
U_i  =  \exp\bigg( i\pi \sum_{j=1}^{i-1}S_j^+S_j^- \bigg).
\end{equation}
These operators now behave like real fermions:
\begin{equation}
\{ f_i^\dagger, f_j\} = \delta_{ij}.
\end{equation}
After performing a Fourier transformation, the Hamiltonian now takes the simple form:
\begin{equation}
H = \sum_{k=1}^N\varepsilon(k) f_k^\dagger f_k,
\end{equation}
where the energy is given by
\begin{equation}
\varepsilon(k) =  4\pi J_\perp \cos \frac{2\pi}{N} k. 
\end{equation}
The form of the spectrum is shown in Figure \ref{fig:spectrum}. We thus find a system of free fermions. If we choose as a ground state of the system the state where all the negative modes are filled, we find that the excitations over this ground state have a linear dispersion relation 
\begin{equation}
\Delta\varepsilon = 4\pi J_\perp \frac{2\pi}{N}\Delta k\equiv v_\text{F} \Delta k. 
\end{equation}

Note that, to derive the above spectrum, we did not assume any lattice spacing. It is enough to know which spins are in relation to each other and how they interact. In this sense the system is background independent. That is, \emph{we do not need to assume a background metric structure to derive the dynamics of the system}. 

How then can we define notions like distance in our system?

The answer lies in the excitation spectrum that we have just derived. A point in the system can be defined by the intersection of two traveling wave-packets made up of the fermionic excitations of equation (\ref{eqn:fermions}). The distance between points is determined once the speed of the traveling wave-packets is \emph{defined}. This is analogous to how we practically define distance, by the value we give to the speed of light $c$. Since the dispersion relation is linear, all wave-packets will travel with the same speed. 

In the limit where the width of the wave-packets comprises of a large number of spins, the discrete spin model will appear smooth to an observer in the model. Because all the excitations have the same speed, the model is then perceived as two-dimensional Minkowski space. Thus, in the background-independent form of this model, the geometry of Minkowski space is contained in the dynamics of the model and not in the kinematics. 

\section{The Cosmological Constant Problem}\label{sec:ccp}
How then would an observer in the model, without access to the microscopic spins, describe the physics of the system? First, she will find that she lives in a two-dimensional Minkowski space. Second, she will find that the particle content can be well described by two-component spinor fields $\psi_\alpha$, $\alpha = 1,2$, with Hamiltonian
\begin{equation}
H = \int dx\ \psi^\dagger(x)\beta i \partial_x\psi(x),
\end{equation}
where
\begin{equation}
\beta = \sigma^1 = \begin{pmatrix}0 & 1\\ 1 & 0\end{pmatrix}.
\end{equation}
The relation of the spinor field $\psi_\alpha$ to the variables $f_i$ of the previous section is given as follows: First we define the two-component spinor $\bar\psi_\alpha$, $\alpha=1,2$:
\begin{eqnarray}
\bar\psi_1(m) & = & i^{2m}\; f(2m) \\
\bar\psi_2(m) & = & i^{2m+1}\; f(2m+1), 
\end{eqnarray}
for $m=1,\ldots,N/2$. The two-component spinor $\psi_\alpha$ is then obtained from $\bar\psi_\alpha$ by appropriate rescaling so that the Fermi velocity $v_\text{F}$ becomes unity (for more details see \cite[chapter 4]{fradkin}). 

\begin{table}
\caption{Fundamental and emergent view.}\label{tab:comparison}
\begin{ruledtabular}
\begin{tabular}{ccc}
 & fundamental view & emergent view \\
 \hline
 \parbox{1cm}{Hilbert space} & $\mathcal{H} = (\mathbb{C}^2)^{\otimes N}$ & \parbox{2cm}{fermionic\\ Fock space $\mathcal{H}^-$}\\
  & & \\
 $\dim \mathcal{H}$ & $2^N$ & $\infty$ \\
  & & \\
\parbox{1cm}{ground state} & \parbox{2cm}{filled Fermi sea $\vert\text{gnd}\rangle$} & \parbox{2cm}{Minkowski space $+$ Fock vacuum $\vert 0 \rangle$}\\
 & & \\
excitations & $f_i$ & $\psi_\alpha$ \\
 & & \\
\parbox{1.8cm}{cosmological\\ constant} & -- & $-\frac{1}{2}\sum_k\hbar\omega_k$. 
\end{tabular}
\end{ruledtabular}
\end{table}

When she quantizes the system, she will do so by constructing a fermionic Fock space $\mathcal{H}^-$. The Fock vacuum $\vert 0 \rangle\in\mathcal{H}^-$ is interpreted by the coarse-grained observer as empty two-dimensional Minkowski space. 

At this point, she runs into the cosmological constant problem. She will notice that all the modes in $\mathcal{H}^-$ contribute $-\frac{1}{2}\hbar\omega$ to the vacuum energy. The contribution to the cosmological constant should thus be
\begin{equation}
-\frac{1}{2}\sum_k\hbar\omega_k.
\end{equation}

The observer's cosmological constant problem is really a paradox that can be resolved if one compares the emergent view, described by the $\psi$'s, with the more fundamental view, described by the $f$'s (see table \ref{tab:comparison}). The Fock vacuum $\vert 0\rangle$ is to be compared to the filled Fermi sea $\vert\text{gnd}\rangle$ and  the $\frac{1}{2}\hbar\omega$ in the $\psi$-description are like the energy of the filled Fermi sea in the $f$-description. The ground state $\vert\text{gnd}\rangle$ does not have any a priori metric properties. It is only when the excitations are examined that the two-dimensional Minkowski space appears, \emph{for excitations over the ground state} $\vert\text{gnd}\rangle$. Thus, the $\frac{1}{2}\hbar\omega$ are \emph{not} amounts of energy sitting on a spacetime and curving it. They are part of a pre-geometry and cannot curve the effective spacetime.

We see that the root of the cosmological constant problem lies in the fact that we have treated as distinct objects which in fact are one and the same. If we treat quantum fields as living on a spacetime, then we will encounter the cosmological constant problem. If, on the other, hand we realize that it is only through the excitations described by the quantum fields that a spacetime appears in the first place, the cosmological constant problem disappears. 

\section{Conclusions}\label{sec:conclusion}
``The world is not given twice". With these words Leibniz tried to convince Clarke, in their correspondence, that it is wrong to view the world as being embedded in a further container \cite{alex}. What he meant was that it is enough to give the relations between the material things making up the world. To add the metric information was like giving the world twice.

In this paper we argued that the cosmological constant problem arises because we describe the world as if it has to be given twice. First, there are the quantum fields, then there is the spacetime on which they live. Using a simple model from solid state physics we showed that no metric information is needed in the initial formulation of the theory, but a two-dimensional Minkowski spacetime can be recovered from the dynamics. This is because the relations between the excitations give the appearance of spacetime in the appropriate limits.

In summary, we constructed a background independent quantum field theory using the simplest model possible. We want to stress though that the resolution of the cosmological constant problem does not depend on the model or its dimensionality. This simple model was sufficient to present the argument, but the goal has to be a model that gives rise to gravity as well as the Standard Model. One expects that such a model will have more complicated couplings. Models that show how fermions and gauge theories arise exist \cite{wen} and fit into our approach. The inclusion of gravity is currently under investigation. 

In our model, we were able to show that the cosmological constant problem is a paradox that appears only if spacetime is regarded as fundamental rather then emergent. Even though we dealt exclusively with a flat spacetime, this is appropriate for the cosmological constant problem since it is usually presented in exactly this context. We argued that the solution of the problem can also be given without the inclusion of gravity. 

In quantum field theory one encounters a number of effects that are due to vacuum fluctuations. Examples include the Casimir effect and the Lamb shift. One might wonder whether these observed effects disappear in our view of quantum field theory. The easiest way to see that this is not a concern is to realize that the description of the low energy physics in terms of a quantum field theory is a valid description in its domain. Low energy phenomena like the Casimir effect or the Lamb shift thus remain untouched. The point of the argument is to show that the quantum field theory picture is just that: a picture. It runs into trouble when applied outside its domain.

Recent experimental data \cite{wmap} indicates that we live in universe with a small cosmological constant. The argument presented here says that there should be no cosmological constant that has its origin in the zero energy of quantum fields. The argument does not exclude other sources for a cosmological constant. 

In our construction, Fock-space-like structures are only approximately present. We can exactly identify one-particle Hilbert spaces but, for interacting theories, we can identify many particle states only when the particles are sufficiently far apart. That Fock spaces are only approximate is also clear from the fact that the dimension of our Hilbert space is finite. It does not have enough room for an infinite-dimensional Fock space. 

A consequence of the setup as it is presented here is that special relativity is no longer fundamental. It only arises in the limit in which the spin model looks smooth and it does so only because the dispersion relation is relativistic. This is to be contrasted with the usual view of quantum field theory in which special relativity plays a fundamental role. In our approach, the emergence of special relativity is a dynamical effect. One striking feature of special relativity is that all particles have the same light cone. This raises an important  question: What dynamical effect ensures that all particles of the theory share the same light cone?

It is important that, while the argument presented in this paper can only be conclusive in a full treatment including gravity, it also indicates where one should look for a quantum mechanical theory that encompasses gravity. It makes clear that any approach to quantum gravity should treat matter and spacetime as two manifestations of the same thing. An approach that puts matter on a spacetime will encounter the cosmological constant problem. 

\begin{acknowledgments}
The author would like to thank J.~Barbour, T.~Jacobson, F.~Markopoulou-Kalamara, D.~Oriti, H.~Pfeiffer, L.~Smolin, J.~Stachel, A.~Valentini, and H.~Westman for fruitful discussions and constructive criticism. 
\end{acknowledgments}

\end{document}